# Progressive Layer Stripping Analysis for HVSR Interpretation

Mersad Fathizadeh[1], Clinton M. Wood[2], Hosna Kianfar[3]

[1]University of Arkansas, Graduate Research Assistant, Dept. of Civil Eng., 4190 Bell Engineering Center Fayetteville, AR 72701, USA, mersadf@uark.edu
[2]University of Arkansas, Associate Professor, Dept. of Civil Eng., 4190 Bell Engineering Center Fayetteville, AR 72701, USA, cmwood@uark.edu
[1]University of Arkansas, Graduate Research Assistant, Dept. of Civil Eng., 4190 Bell Engineering Center Fayetteville, AR 72701, USA, hkianfar@uark.edu

**ABSTRACT**

The horizontal-to-vertical spectral ratio (HVSR) technique is widely used to determine site fundamental periods from ambient noise recordings, but relating the observed peak to a specific impedance contrast within layered soils remains challenging. This paper presents an enhanced implementation of hvstrip-progressive, a Python package for forward HVSR modelling under the diffuse-field assumption and systematic progressive layer stripping. The package computes theoretical HVSR curves from shear-wave velocity (Vs) profiles, iteratively removes the deepest finite layer and promotes the next layer to half-space, and tracks how the fundamental frequency and amplitude change with each step. Compared with previous implementations, the software now supports adaptive frequency scanning, rigorous model validation, and publication-quality visualizations. Using a synthetic seven-layer soil profile, we show that the fundamental peak shifts from 6.99 Hz to 23.45 Hz as layers are stripped and that the maximum impedance contrast of 1.46 at 17 m depth controls the resonance. The transparent workflow, reproducible outputs and open-source distribution make hvstrip-progressive a practical tool for seismic site characterization and microzonation studies.

# 1 INTRODUCTION

## 1.1 Background and motivation

Near-surface velocity structure exerts fundamental control over seismic ground motion characteristics, necessitating robust characterization methods for accurate hazard assessment. The horizontal-to-vertical spectral ratio (HVSR) technique, pioneered by Nakamura (1989), leverages ambient noise recordings to determine site-specific resonant frequencies through straightforward spectral analysis. This approach has gained widespread adoption within the geophysical community, particularly for microzonation initiatives and landslide susceptibility assessments, owing to its operational simplicity and minimal site disturbance requirements. Contemporary investigations further underscore the method's utility when integrated into comprehensive geophysical campaigns, as evidenced by Rahimi et al. (2025), who demonstrated enhanced subsurface resolution through strategic multi-method deployment at structurally complex sites.

The theoretical underpinnings of HVSR peak generation remain subject to considerable scientific discourse. Initial interpretations attributed peak characteristics either to elliptical particle motion associated with fundamental-mode Rayleigh waves or to shear-wave resonance phenomena within unconsolidated sediments. Contemporary theoretical frameworks, grounded in diffuse-field analysis, establish that HVSR measurements fundamentally represent the square root of ratios between horizontal and vertical Green's function imaginary components—a formulation that inherently incorporates contributions from the complete wavefield spectrum. This theoretical advancement enables direct HVSR computation from stratified velocity models without requiring explicit wave-type decomposition or restrictive assumptions regarding dominant propagation modes.

## 1.2 Interpretation challenges

Multilayered geological environments present substantial interpretive complexities, wherein multiple velocity contrasts generate overlapping resonance signatures across similar frequency ranges. Conventional processing workflows typically isolate primary frequency peaks from HVSR curves; however, establishing definitive correlations between observed peaks and specific subsurface interfaces remains problematic. Complete waveform inversion of HVSR datasets suffers from inherent non-uniqueness and computational inefficiency, particularly for sites with complex stratigraphy. While joint inversion schemes incorporating surface wave dispersion data or empirical transfer functions demonstrate improved parameter resolution, such approaches necessitate supplementary field measurements that may compromise the method's principal advantage of operational efficiency. These limitations underscore the critical need for developing streamlined, reproducible analytical frameworks capable of establishing quantitative relationships between HVSR spectral characteristics and discrete subsurface velocity contrasts.

### 1.3 Progressive layer stripping

To address this challenge, we employ **progressive layer stripping**, an algorithm that iteratively removes the deepest finite layer from a velocity model and promotes the next layer to half-space. At each step, the HVSR is recomputed, and the fundamental frequency, amplitude and impedance contrast are recorded. By tracking how the peak evolves, the method identifies which interface controls the observed resonance. This approach provides a transparent and physically interpretable complement to inversion and transfer-function methods.

### 1.4 Contributions

This paper builds on a previous implementation of **hvstrip-progressive** by introducing the following enhancements:

1. **Adaptive frequency scanning** to ensure peaks are captured across a wide range of models and to avoid aliasing at band edges. The scanning adjusts the frequency band if the peak lies near the limits[1].

2. **Rigorous model validation** that checks physical consistency (e.g., $V_p > V_s$, positive densities, unique half-space) before computation.

3. **Publication-quality visualizations** including waterfall plots, peak evolution graphs and multi-panel summary figures.

4. **Comprehensive reporting** that outputs CSV/JSON tables, figures and an analysis report summarizing peak shifts and impedance contrasts.

The remainder of this paper summarizes the theoretical framework, details the progressive stripping algorithm and software architecture, presents an illustrative example, and discusses the practical implications of this methodology.

## 2 METHODOLOGY AND DATA PROCESSING

**Theoretical framework**

### 2.1 HVSR under the diffuse-field assumption

Ambient noise can be modelled as a diffuse field—a superposition of body and surface waves travelling in all directions with uniformly distributed energy. Under this assumption, Sánchez-Sesma et al. (2011) showed that the HVSR is given by

$$H/V(\omega) = \sqrt{\frac{I_m[G_{11}(\omega)] + I_m[G_{22}(\omega)]}{I_m[G_{33}(\omega)]}},$$

where $G_{ij}(\omega)$ are Green's function components for displacement in direction $i$ due to a unit force in direction $j$ and $I_m[\cdot]$ denotes the imaginary part. Indices 1 and 2 correspond to horizontal directions and 3 to the vertical. This expression shows that all elastic wave types contribute to the HVSR and enables direct forward calculation from a layered model.

**2.2 Forward HVSR calculation**

For horizontally stratified media, Green's functions are computed via wavenumber integration and propagator matrix techniques. Each layer is characterized by thickness $h_i$, P-wave velocity $V_{p,i}$, S-wave velocity $V_{s,i}$ and density $\rho_i$. The HVf solver (García-Jerez et al., 2016) separates contributions of Rayleigh, Love and body waves and integrates them over horizontal wavenumber to form the full Green's functions. Numerical stability at high frequency is ensured through careful treatment of exponentially growing terms. In hvstrip-progressive, the solver is wrapped in Python so that theoretical HVSR curves can be calculated for arbitrary models.

**2.3 Progressive layer stripping algorithm**

Given an initial model with $N$ finite layers over a half-space, the progressive stripping algorithm proceeds as follows:

**1. Base model computation**: Compute the HVSR curve and identify the fundamental peak frequency $f_0$ and amplitude $A_0$.

2. **Layer removal**: Remove the deepest finite layer and promote the next finite layer to half-space by setting its thickness to zero while retaining its $V_p$, $V_s$ and $\rho$. This ensures there is only one half-space and avoids duplication of properties.

3. **Recomputation**: Compute the new HVSR curve and identify the updated peak frequency and amplitude.

4. **Recording**: Calculate the impedance contrast $IC = \frac{\rho_{below} V_{s,below}}{\rho_{above} V_{s,above}}$ at the removed interface and record the frequency shift relative to the base model.

5. **Iteration**: Repeat steps 2–4 until only the half-space remains.

The algorithm is summarized in pseudocode in Algorithm 1 (see previous work). It ensures a consistent half-space at each step and provides a sequence of peak frequencies and impedance contrasts for interpretation.

## 2.4 Peak detection and impedance contrast

Peak identification combines derivative-based detection and prominence filtering. Local maxima are found and those with amplitudes less than 20 % of neighboring minima are discarded. The fundamental peak is the lowest-frequency peak satisfying these criteria. To avoid aliasing, the frequency band is expanded if the peak lies near the band edges[1].

Impedance contrasts quantify the strength of interfaces and are calculated as $IC = \frac{Z_{below}}{Z_{above}}$, where seismic impedance $Z = \rho V_s$ . High contrasts correspond to strong reflections and large HVSR peaks. Tracking how $IC$ changes with depth helps identify the controlling interface.

## 3 Software implementation

### 3.1 Package structure and validation

The hvstrip-progressive package adopts a modular structure with separate modules for forward calculation, layer stripping, post-processing, plotting and reporting. Model validation routines check physical constraints before computation: positive layer thicknesses, positive densities, $V_p > V_s$ , and a single half-space. These checks prevent numerical instabilities and ensure meaningful results.

### 3.2 Adaptive frequency scanning

To capture resonance peaks across diverse conditions, the package implements adaptive scanning. If the detected peak lies within 10 % of the lower or upper frequency limit, the band is expanded (halved or doubled) and the HVSR recomputed. This process repeats for up to three passes, ensuring that peaks are not missed due to an insufficient frequency range[1].

### 3.3 Visualization and reporting

The visualization module produces multiple plot types:

- **Individual step plots**: Vs profiles and HVSR curves with peak annotations (Figures 1–2).

- **Peak evolution analysis**: Combined line and bar charts showing how peak frequency, amplitude and relative shift change with stripping step (Figure 3).

- **Curve overlay**: Waterfall plot overlaying all HVSR curves to visualize the continuous migration of the resonance (Figure 4).

- **Multi-panel summary**: A publication-ready figure summarizing HVSR evolution, peak evolution, impedance contrast profiles and key numeric results (Figure 5).

The reporting module collates all results—tables, figures and parameter settings—into a structured analysis report (see the Appendix of the provided report). Outputs are saved in both CSV/JSON formats and high-resolution images suitable for inclusion in manuscripts.

## 3 RESULTS

**Application example: seven-layer soil profile**

### 4.1 Soil profile description

We demonstrate the methodology using a synthetic seven-layer profile representative of soft sediments over stiff bedrock. The layer parameters (Table 1) are derived from the Soil_Profile_Model.txt supplied with the package. The model transitions from low velocities and densities near the surface to high velocities in the half-space, producing several impedance contrasts. The base Vs profile and corresponding theoretical HVSR curve are shown in Figure 1.

In Table 1., Depth to bottom is cumulative thickness; impedance is computed as $Z=\rho V_s$ .

**Table 1**: Soil model used for progressive layer stripping.

| Layer | Thickness (m) | $V_p$ (m/s) | $V_s$ (m/s) | Density (kg/m³) | Depth to bottom (m) | Impedance (kg/m²·s) |
|---|---|---|---|---|---|---|
| 1 | 3 | 685 | 345 | 1586 | 3 | $5.47 \times 10^5$ |
| 2 | 4 | 870 | 438 | 1683 | 7 | $7.37 \times 10^5$ |
| 3 | 4 | 994 | 501 | 1741 | 11 | $8.72 \times 10^5$ |
| 4 | 6 | 1263 | 636 | 1848 | 17 | $1.18 \times 10^6$ |
| 5 | 8 | 1534 | 885 | 1940 | 25 | $1.72 \times 10^6$ |
| 6 | 11 | 2002 | 1156 | 2074 | 36 | $2.40 \times 10^6$ |
| 7 | 13 | 2230 | 1288 | 2130 | 49 | $2.74 \times 10^6$ |
| Half-space | 0 | 2566 | 1482 | 2206 | - | $3.27 \times 10^6$ |

### 4.2 Results of progressive layer stripping

The analysis produced seven stripping steps (0–6). Selected results are illustrated below.

#### 4.2.1 Base model and first stripping step

Figure 1a displays the initial Vs profile and the base theoretical HVSR curve. The base model exhibits a fundamental peak at $f_0$=6.99 Hz with amplitude $A_0$=6.3 . The deepest finite layer has $V_s$=1288 m/s, which transitions to the half-space with $V_s$=1482 m/s and introduces a moderate impedance contrast.

Removing the deepest finite layer yields the first stripped model. Figure 1b shows the Vs profile after stripping and indicates the new fundamental frequency $f_1$=7.84 Hz. The increase reflects the loss of the thick, stiff basal layer and demonstrates how resonance shifts toward higher frequencies as the model becomes shallower.

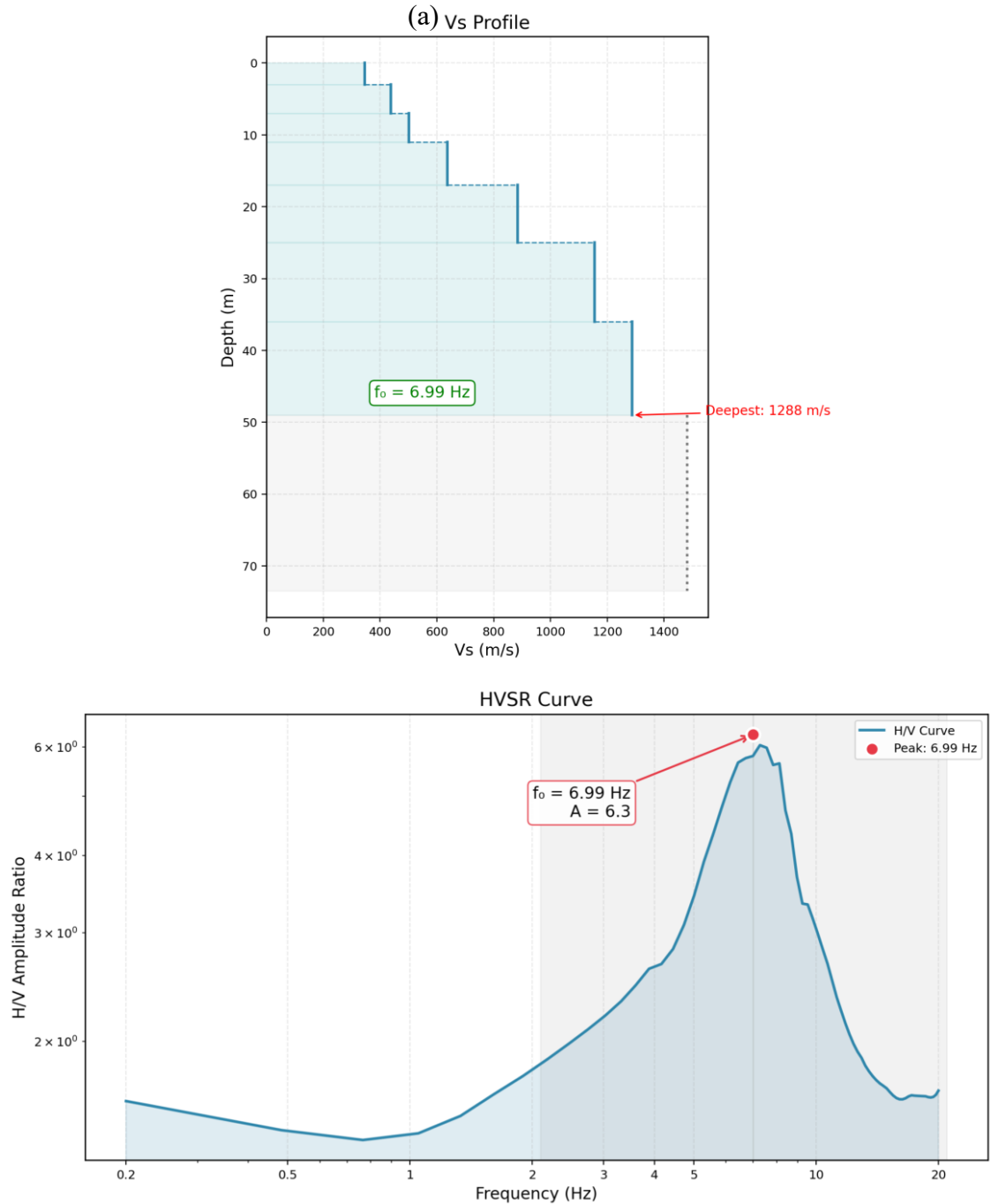

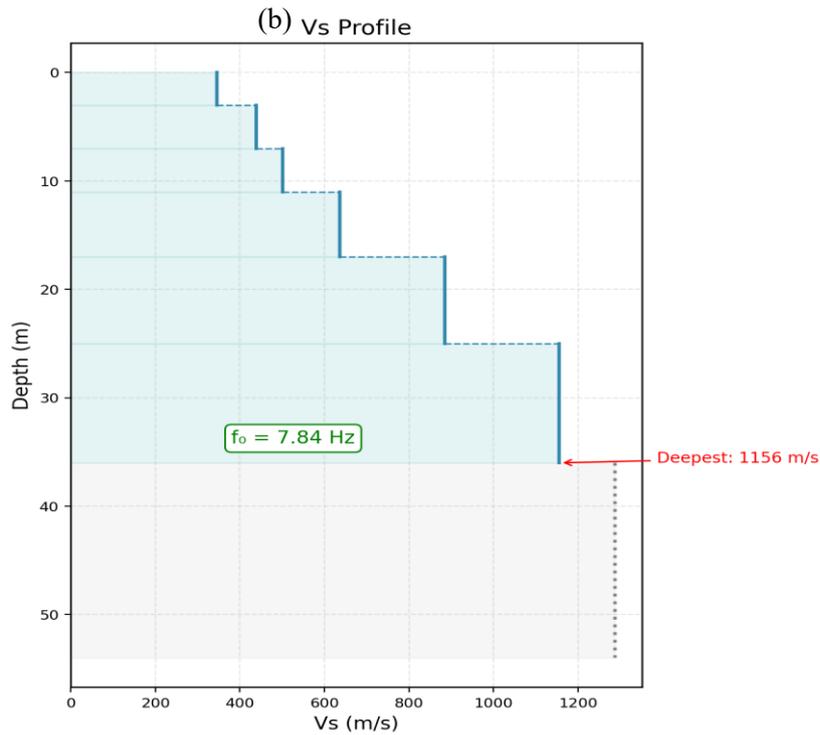

**Figure 1**: (a) Vs profile and HVSR curve for the base model (Step 0) and (b) Vs profile after the first stripping step. The base profile exhibits a fundamental frequency of 6.99 Hz, while the first stripped profile shows a higher resonance at 7.84 Hz.

### 4.2.2 Peak evolution analysis

Figure 2 summarizes how peak frequency, amplitude and relative frequency shift evolve with stripping step. The peak frequency increases gradually from 6.99 Hz to 8.69 Hz over the first three steps, remains near 8–9 Hz for the middle steps, then jumps sharply to 14.89 Hz at step 5 and to 23.45 Hz at the final step. Amplitude decreases monotonically, reflecting the reduction of impedance contrast as layers are removed. The total frequency shift relative to the base model is 235.5 %.

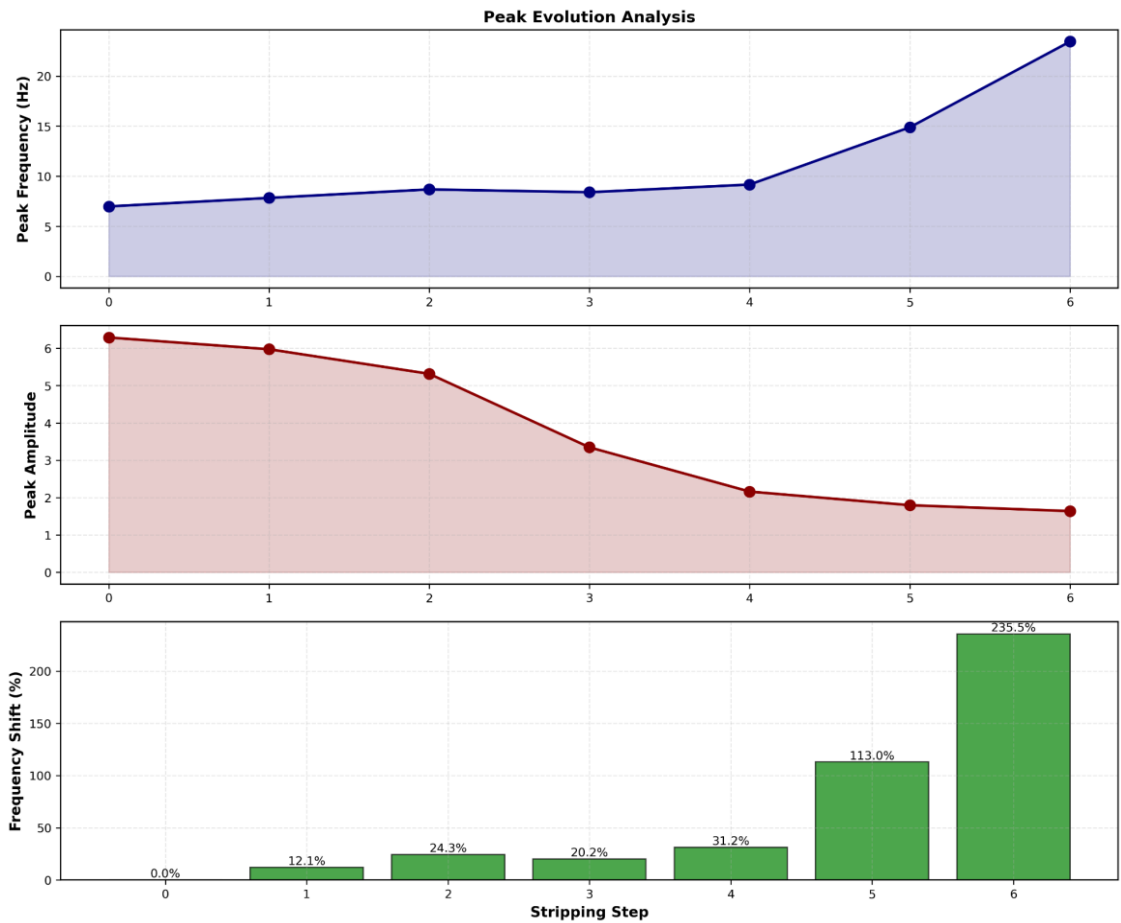

**Figure 2**: Peak evolution analysis summarizing peak frequency (top), peak amplitude (middle) and relative frequency shift (bottom) versus stripping step.

### 4.2.3 HVSR curve overlay

Figure 3 shows all HVSR curves overlaid on a single axis, highlighting the continuous migration of the main resonance. The waterfall plot reveals how secondary peaks become more pronounced as deeper layers are stripped, while the dominant peak moves to higher frequency and decreases in amplitude.

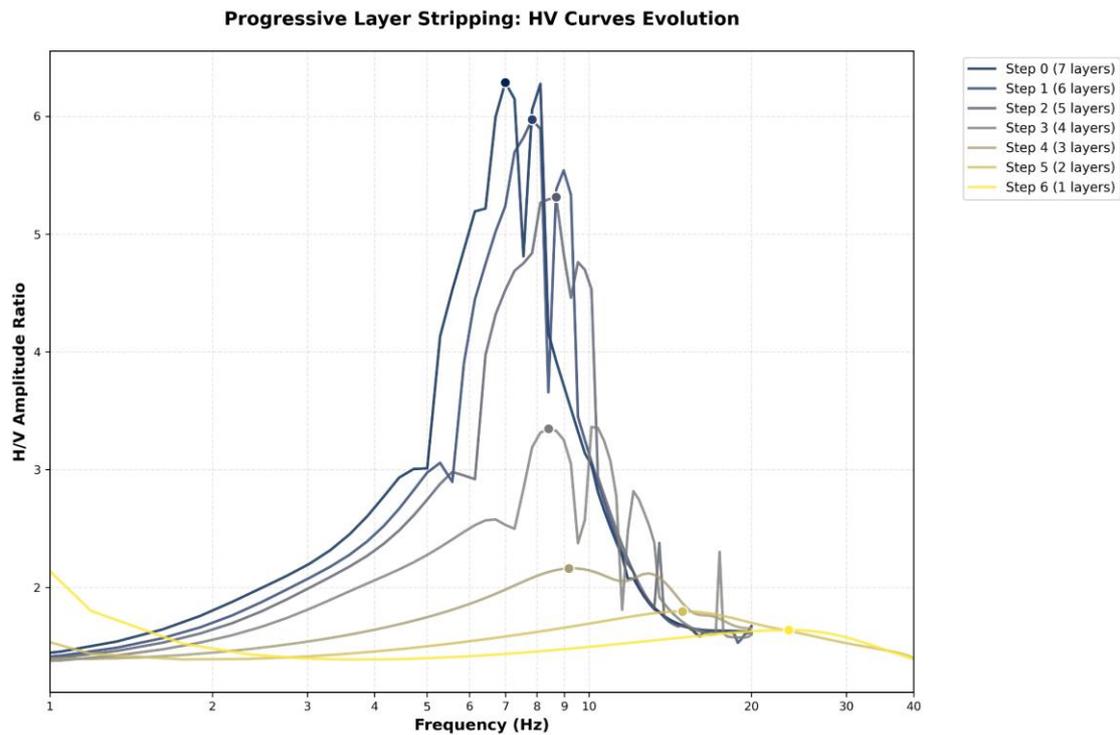

**Figure 3**: HVSR curves for all stripping steps overlaid on a single axis. Darker curves correspond to models with more layers (Step 0), and lighter curves correspond to models with fewer layers.

### 4.2.4 Multi-panel summary

Figure 4 combines key results into a four-panel layout: (a) HVSR evolution for selected steps (0, 3, 6), (b) peak frequency evolution, (c) impedance contrast profile versus depth, and (d) a table of key numeric results. The maximum impedance contrast of 1.46 occurs at the 17 m interface and aligns with the controlling resonance. This figure can be inserted directly into conference papers or reports.

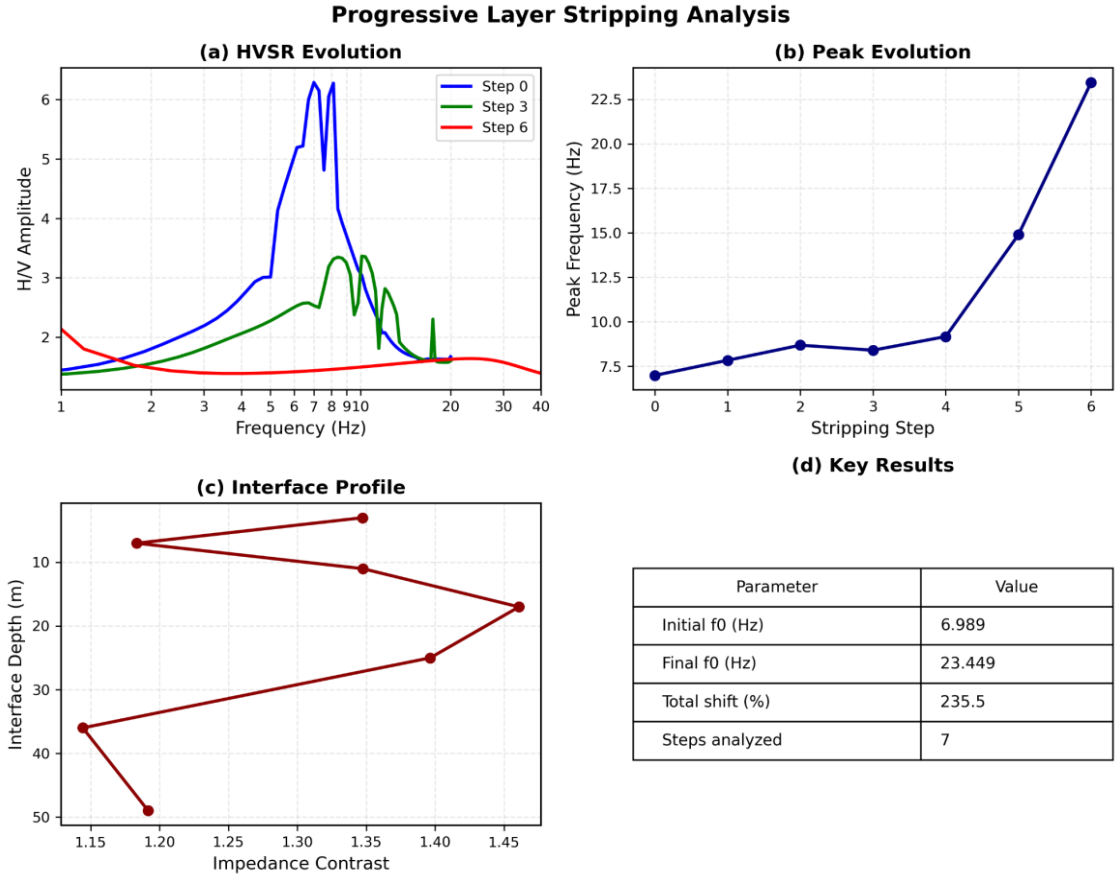

**Figure 4**: Multi-panel summary of the progressive stripping analysis: (a) HVSR evolution (steps 0, 3, 6), (b) peak frequency evolution, (c) impedance contrast versus depth, and (d) key numeric results.

### 4.3 Interpretation

The progressive layer stripping analysis shows that the fundamental peak at 6.99 Hz originates from the impedance contrast at 17 m depth, not from the stronger contrasts at greater depths. This conclusion is consistent with the quarter-wavelength approximation, which predicts the resonant depth as $\frac{\lambda}{4} = \frac{V_s^{avg}}{4f_0}$, yielding approximately 21 m. The largest frequency jumps occur after steps 4 and 5, indicating a transition between resonance modes. The monotonic decrease in amplitude implies that deeper layers contribute more strongly to energy trapping. Collectively, these observations illustrate the value of progressive stripping for identifying the controlling interface and understanding resonance mechanisms.

## 4 DISCUSSION

### 5.1 Advantages of progressive stripping

The method offers several advantages:

5. **Physical transparency**: By analyzing how the HVSR changes when specific layers are removed, one can directly link resonances to individual interfaces.

6. **Reduced ambiguity**: Unlike inversion algorithms that may yield non-unique solutions, progressive stripping provides deterministic insights into resonance control.

7. **Complementarity**: The method can validate or guide inversion results and help design targeted field investigations.

8. **Practicality**: Implementation in Python with a CLI interface makes the workflow reproducible and accessible.

### 5.2 Limitations and future work

Important limitations include the assumption of 1-D stratigraphy and purely elastic behavior. In areas with lateral heterogeneity or strong damping, the results may not capture 3-D effects or attenuation. The algorithm also requires a pre-existing velocity model; its utility therefore depends on the availability and quality of Vs profiles. Future work could extend the method to account for frequency-dependent Q, include stochastic uncertainty quantification, and integrate machine learning for automated peak classification.

### 5.3 Engineering applications

Progressive stripping is particularly useful for site response studies, microzonation and preliminary design. It can identify the depth of the controlling impedance contrast, inform borehole or geophysical survey planning, and calibrate Vs30 proxies. Because the method uses only a velocity model and no additional data, it is well suited to preliminary assessments or resource-constrained projects.

## 5 CONCLUSION

We have presented an enhanced implementation of hvstrip-progressive for forward HVSR modelling and progressive layer stripping. The package provides adaptive frequency scanning, rigorous model validation, high-quality visualizations and comprehensive reporting. Applied to a seven-layer profile, the method demonstrated a 235 % shift in fundamental frequency and identified the 17 m interface as the resonance controller. The workflow is reproducible, transparent and open source, making it a valuable addition to the seismological toolkit.

---

[1] GitHub

https://github.com/mersadfathizadeh1995/hvstrip-progressive/blob/dcd95499649b194d89fb8fb6bd3fd6951bfb54f1/hvstrip_progressive/core/report_generator.py